\documentclass[aps,prl,twocolumn,groupedaddress,showpacs]{revtex4}
\usepackage{amsfonts}
\usepackage{mathrsfs}
\usepackage{graphicx}
\usepackage{amsmath}
\usepackage{amssymb}
\usepackage{amsbsy}

\begin{document}

\title{Advanced Mean Field Theory of the Restricted Boltzmann Machine}

\author{Haiping Huang}
\email{physhuang@gmail.com; sites.google.com/site/physhuang}
\address{RIKEN Brain Science Institute, Wako-shi, Saitama
351-0198, Japan}
\author{Taro Toyoizumi}
\affiliation{RIKEN Brain Science Institute, Wako-shi, Saitama
351-0198, Japan}
\date{\today}

\begin{abstract}
Learning in restricted Boltzmann machine is typically hard due to
the computation of gradients of log-likelihood function. To describe
the network state statistics of the restricted Boltzmann machine, we
develop an advanced mean field theory based on the Bethe
approximation. Our theory provides an efficient message passing
based method that evaluates not only the partition function (free
energy) but also its gradients without requiring statistical
sampling. The results are compared with those obtained by the
computationally expensive sampling based method.

\end{abstract}

 \maketitle

Restricted Boltzmann machine (RBM) forms building blocks of a deep
belief network~\cite{Hinton-2006a,Bengio-2013}, which is able to
learn complex internal representations of structured objects (such
as nature image, speech, or hand writing). RBM also has wide
applications in computational biology problem, for example, modeling
high-dimensional neural activity data from cortical microcolumns
~\cite{koster-2014}.

However, learning in RBM is computationally hard, since gradients of
the log-likelihood function needs to be computed at every iteration
step to update the model parameters. This computation is usually
accomplished by Gibbs-sampling-based method or its
variants~\cite{Hinton-2002,Hinton-2006b}, for which the tradeoff
between accuracy and convergence speed requires careful
considerations. Furthermore, an efficient way to evaluate the
partition function (e.g., log-likelihood function for
cross-validation analysis) remains unknown.

Here, we develop a mean field theory for the RBM based on the cavity
method (Bethe approximation)~\cite{cavity-2001}, which yields an
efficient and fully-distributed algorithm to evaluate the
free-energy (partition function) of a RBM of interest. The
remarkable efficiency is confirmed by comparing the computation
results of gradients of log-likelihood function by Gibbs sampling
and the proposed mean field theory.

\begin{figure}
\centering
    \includegraphics[bb=16 566 532 668,scale=0.45]{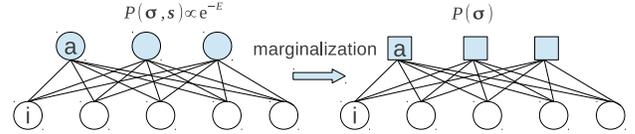}
  \caption{ (Color online) A RBM is composed of one hidden layer and one visible layer. No lateral connections exist within both
  hidden and visible layers. Each hidden node is connected to all visible nodes with symmetric coupling weights, and is responsible for capturing high order dependence.
  The original RBM is shown in the left panel as an example of three hidden nodes (solid circles) and five visible nodes (empty circles).
  The right panel shows a transformed factor graph after marginalization of hidden states
    for theoretical analysis. Each factor node (square node) represents the probabilistic normalization
  of a hidden node given the state of all visible nodes (see the main text).
  }\label{rbm}
\end{figure}

A RBM~\cite{Smolensky-1986,Freund-1994} consists of one hidden layer
and one visible layer without lateral connections between nodes in
each layer. We assume the hidden layer has $M$ nodes, while the
visible layer has $N$ nodes. Hidden node $a$ with external field
$h_a$ is connected to visible node $j$ with field $\phi_j$ by a
symmetric coupling $w_{aj}$. The energy function for RBM is thus
defined by $E=-\sum_{i,a}\sigma_iw_{ai}s_a-\sum_i\sigma_i\phi_i
-\sum_ah_as_a$, where $\sigma_i$ and $s_a$ are used to specify the
state of visible node $i$ and hidden node $a$, respectively. Due to
the conditional independence of hidden nodes' state given
$\boldsymbol{\sigma}$, the state of the hidden nodes can be
marginalized. This leads to the following probability of a visible
state:
\begin{equation}\label{Pobs}
 P(\boldsymbol{\sigma})=\frac{1}{Z}\prod_{a}\left[2\cosh(\boldsymbol{w}_a\cdot\boldsymbol{\sigma}+h_a)\right]\prod_{i}e^{\phi_i\sigma_i},
\end{equation}
where $\boldsymbol{w}_a$ denotes the $a$-th row vector of the
coupling matrix $\boldsymbol{w}$. $Z$ is a normalization constant
(also called the partition function) of the model. As a model study,
we assume the element of the matrix $\boldsymbol{w}$ is
independently and identically distributed with a normal distribution
with mean zero and variance $g/N$. We assume that the external field
for both layers follows a normal distribution with mean zero and
variance $v$. We denote the ratio between the number of hidden nodes
and that of visible nodes by $\alpha=M/N$, where $M$ and $N$ can be
arbitrarily large. A schematic representation of a RBM ($M=3$,
$N=5$) is shown in Fig.~\ref{rbm}.

An exact computation of $Z$ requires an exponential computational
complexity ($2^N$), which becomes impossible for a relatively large
$N$. However, advanced mean field approximation can be used to
compute approximate values under certain condition, and its
prediction should be compared with numerical simulations. Here, we
propose the Bethe approximation~\cite{cavity-2001} to tackle this
problem. First, we transform the original model (left panel of
Fig.~\ref{rbm}) into a factor graph (right panel of
Fig.~\ref{rbm})~\cite{Frey-2001}, where each square node indicates a
Boltzmann factor
$2\cosh(\boldsymbol{w}_a\cdot\boldsymbol{\sigma}+h_a)$ in
Eq.~(\ref{Pobs}). Then, we introduce the cavity probability
$P_{i\rightarrow a}(\sigma_i)$ that the visible node $i$ takes state
$\sigma_i$ in the absence of the contribution from factor node
$a$~\cite{MM-2009}, and $P_{i\rightarrow a}(\sigma_i)$ satisfies the
following self-consistent equations:
\begin{subequations}\label{bp0}
\begin{align}
P_{i\rightarrow a}(\sigma_{i})&\propto
e^{\phi_i\sigma_{i}}\prod_{b\in\partial i\backslash
a}\mu_{b\rightarrow i}(\sigma_{i}),\label{bp2}\\
\begin{split}
\mu_{b\rightarrow i}(\sigma_{i})&=\sum_{\{\sigma_{j}|j\in\partial
  b\backslash
  i\}}2\cosh\left(\boldsymbol{w}_{b}\cdot\boldsymbol{\sigma}+h_b\right)\prod_{j\in\partial
  b\backslash i}P_{j\rightarrow b}(\sigma_{j}),\label{bp1}
\end{split}
\end{align}
\end{subequations}
where the symbol $\propto$ indicates a normalization constant,
$\partial i\backslash a$ denotes the neighbors of node $i$ except
factor node $a$, $\partial b\backslash i$ denotes the neighbors of
factor node $b$ except visible node $i$, and the auxiliary quantity
$\mu_{b\rightarrow i}(\sigma_i)$ represents the contribution from
factor node $b$ to visible node $i$ given the value of
$\sigma_i$~\cite{MM-2009}. With these definitions, the products in
Eq.~(\ref{bp0}) are reasonable under the weak correlations
assumption, whereas, the validity of this Bethe approximation should
be checked by a stability analysis.

Note that the computation in Eq.~(\ref{bp1}) is still intractable
due to the summation over all possible $\boldsymbol{\sigma}$ except
$\sigma_i$. However, because $\mathcal {U}_{b\rightarrow
i}\equiv\sum_{j\in\partial b\backslash i}w_{bj}\sigma_j$ is a sum of
a large number of nearly independent random variables, the
central-limit theorem implies that the distribution of $\mathcal
{U}_{b\rightarrow i}$ is well characterized by its mean and
variance~\cite{Huang-JPA2013}, i.e., $G_{b\rightarrow
i}=\sum_{j\in\partial b\backslash i}w_{bj}m_{j\rightarrow b}$ and
$\Xi^{2}_{b\rightarrow i}\simeq\sum_{j\in\partial b\backslash
i}w_{bj}^{2}(1-m_{j\rightarrow b}^{2})$ respectively, where
$m_{j\rightarrow b}\equiv \sum_{\sigma_j}\sigma_j P_{j\rightarrow
b}(\sigma_j) $ denotes the cavity magnetization (the average of
$\sigma_j$ in the absence of factor node $b$).

Because we consider the binary spin variable $\sigma_i=\pm1$,
$P_{i\rightarrow a}(\sigma_i)$ and $\mu_{b\rightarrow i}(\sigma_i)$
can be parametrized by $m_{i\rightarrow a}$ and cavity bias
$u_{b\rightarrow i}$, respectively. $u_{b\rightarrow i}$ is defined
as $\frac{1}{2}\ln\frac{\mu_{b\rightarrow
i}(\sigma_i=1)}{\mu_{b\rightarrow i}(\sigma_i=-1)}$. The practical
recursive equations, the so-called message passing equations, are
thus derived as:
\begin{subequations}\label{bp3}
\begin{align}
m_{i\rightarrow a}&=\tanh\left(\phi_i+\sum_{b\in\partial i\backslash
a}u_{b\rightarrow i}\right),\\
u_{b\rightarrow i}&=\frac{1}{2}\ln\frac{\cosh(h_b+G_{b\rightarrow
i}+w_{bi})}{\cosh(h_b+G_{b\rightarrow i}-w_{bi})},\label{bp4}
\end{align}
\end{subequations}
where the $\Xi_{b\rightarrow i}$ dependency in Eq.~(\ref{bp4}) drops
because of the symmetry of $\cosh$. The cavity magnetization can be
understood as the message passing from the visible node to the
factor node, while the cavity bias is interpreted as the message
passing from the factor node to the visible node. This message
passing based computation is much more accurate than naive mean
field approximation~\cite{Kappen-1998}, which assumes a fully
factorized distribution for Eq.~(\ref{Pobs}). In contrast,
Eq.~(\ref{bp3}) captures nearest neighbors' correlations.

Once the iteration of Eq.~(\ref{bp3}) converges, the free energy of
the model can be computed from the fixed-point solution. Under the
Bethe approximation, the Bethe free energy is expressed
as~\cite{Huang-JPA2013,MM-2009}:
\begin{equation}\label{fre}
    F=-\sum_{i}\ln Z_i+(N-1)\sum_{a}\ln Z_a,
\end{equation}
where $Z_i=e^{\phi_i}\prod_{b\in\partial i}\mu_{b\rightarrow
i}(+1)+e^{-\phi_i}\prod_{b\in\partial i}\mu_{b\rightarrow i}(-1)$,
in which $\mu_{b\rightarrow i}(\sigma_i)=2e^{\Xi^{2}_{b\rightarrow
i}/2}\cosh(h_b+G_{b\rightarrow i}+w_{bi}\sigma_i)$.
$Z_a=2e^{\Xi^{2}_{a}/2}\cosh(h_a+G_{a})$. $G_a$ and $\Xi^{2}_{a}$
are given by $\sum_{j\in\partial a}w_{aj}m_{j\rightarrow a}$ and
$\sum_{j\in\partial a}w_{aj}^{2}(1-m_{j\rightarrow a}^{2})$
respectively. Each stable solution of the message passing algorithm
of Eq.~(\ref{bp3}) corresponds to a local minimum of the free energy
function in Eq.~(\ref{fre})~\cite{Heskes-2004}.

RBM defined in Fig.~\ref{rbm} is basically a densely-connected
graphical model. Our mean field theory provides a practical way to
estimate the free energy of {\em single instances} (typical examples
of the model). More precisely, we initialize the cavity
magnetization and bias on each link of the factor graph by random
values, and then iterate Eq.~(\ref{bp3}) until it converges within a
prescribed accuracy. Note that the overall time complexity is of the
order $\mathcal {O}(N^2)$, furthermore, the algorithm is
fully-distributed, and thus amenable to large-scale applications.


\begin{figure}
          \includegraphics[bb=63 21 740 538,scale=0.35]{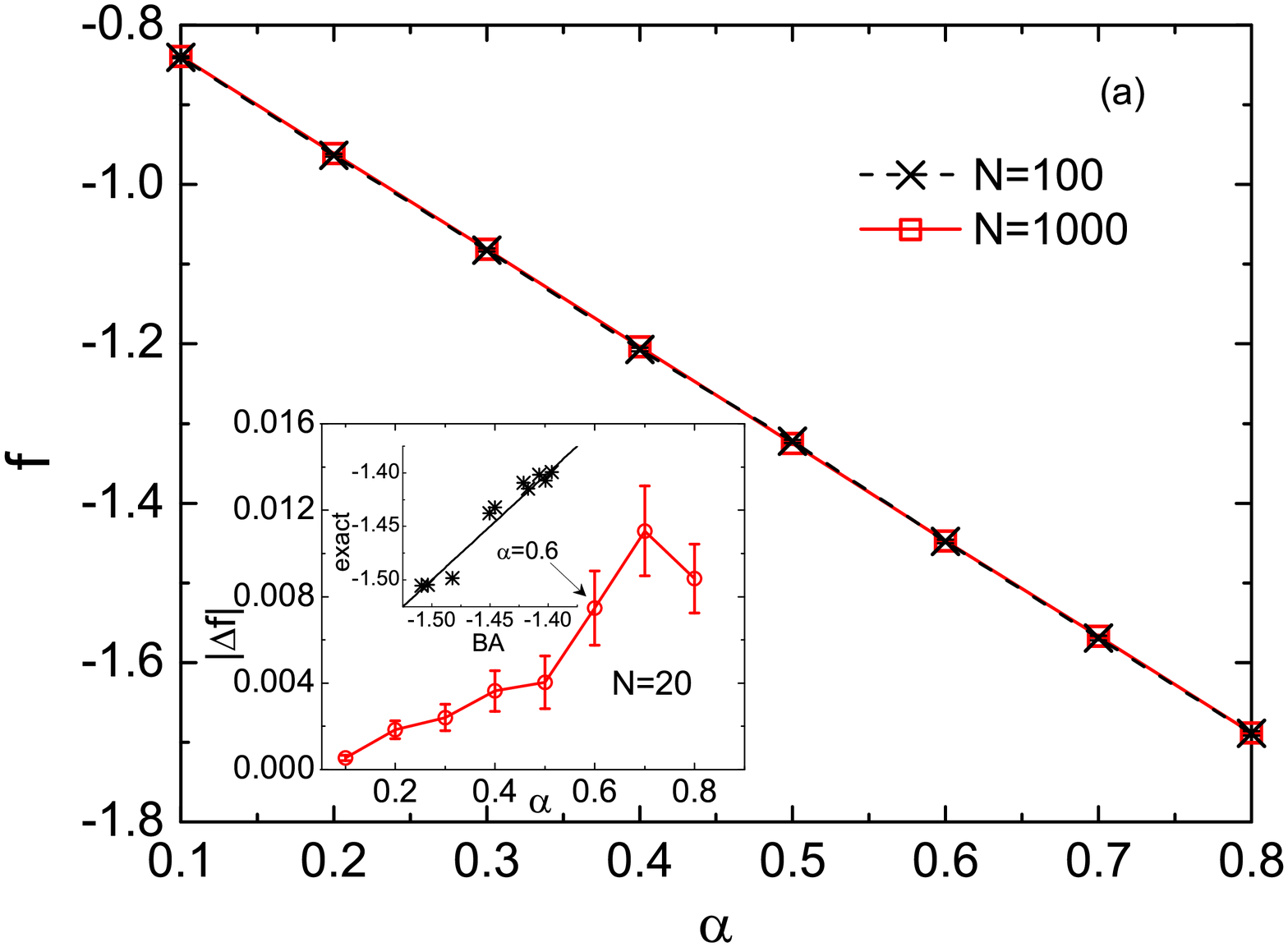}
     \vskip .1cm
     \includegraphics[bb=53 14 742 539,scale=0.35]{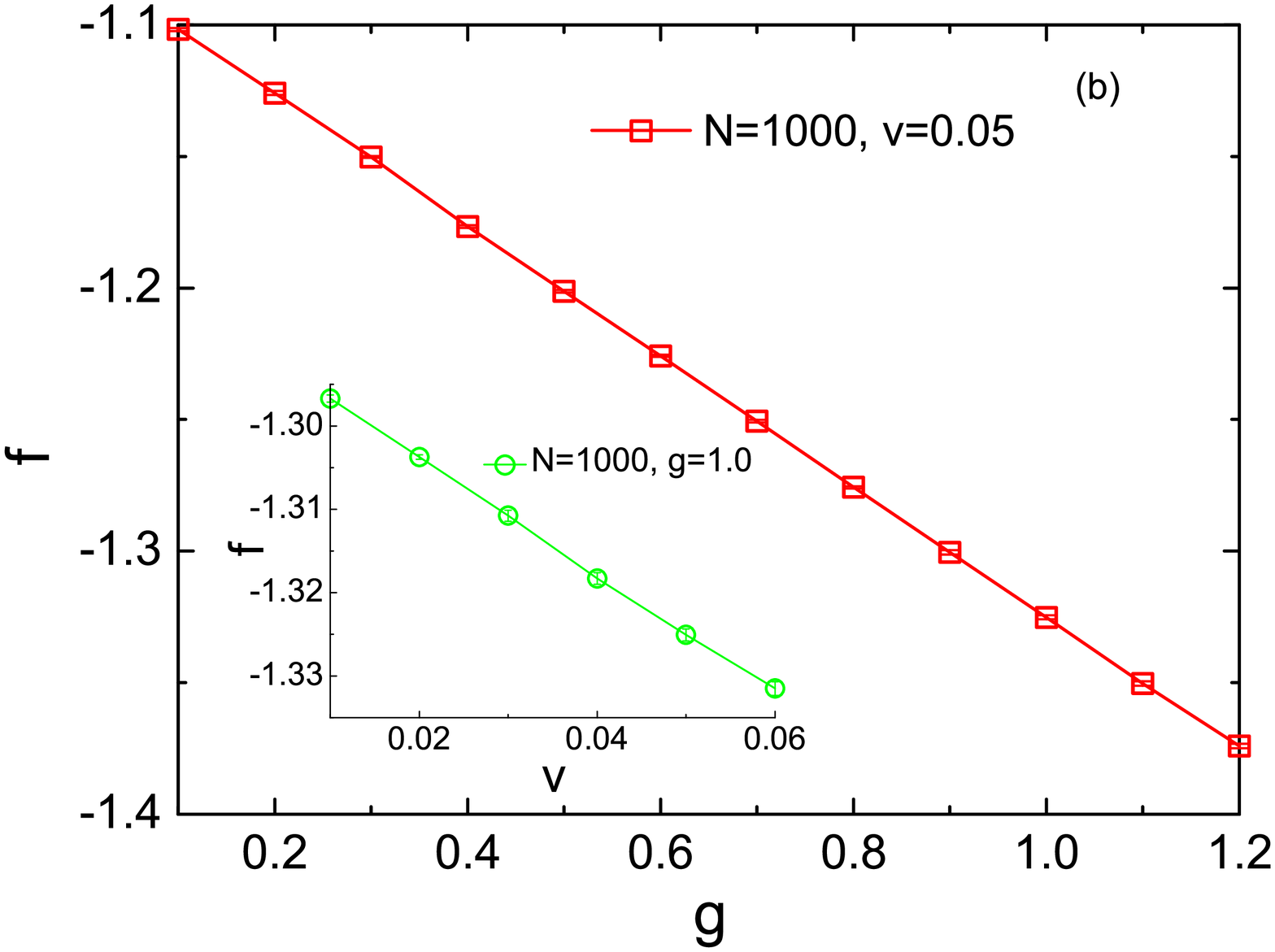}
 \vskip .1cm
  \caption{(Color online)
     Free energy density ($f=F/N$) of single instances of RBMs. Iterations of the
     message passing
      equations are always converged to produce the data points. The error bars give statistical errors
      across ten random instances. (a) free energy density as a function
of $\alpha$ (density of hidden nodes). $g=1.0$, and $v=0.05$. The
inset shows the absolute difference, $|\Delta f|$, of the free
energy density estimated by the exact enumeration and the Bethe
approximation (BA) for $N=20$ (see a comparison for ten instances
with $\alpha=0.6$ in the inset, where the line indicates equality).
      (b) free energy density as a function of weight strength $g$ and field strength $v$ (shown in the inset). $\alpha=0.5$. }\label{FreeE}
 \end{figure}

In the remaining part, we demonstrate the computation of the free
energy on large single instances by applying the message passing
algorithm, and confirm the accuracy of the results by comparing
gradients of the log-likelihood with those obtained by the Gibbs
sampling method. A stability analysis of the message passing
algorithm is also presented.


We run the message passing equations on single instances of RBMs as
size $N$, hidden-node density $\alpha$, and coupling strength $g$
are varied. As displayed in Fig.~\ref{FreeE} (a), the free energy
density decreases as $\alpha$ increases. Note that the density does
not change significantly at two large sizes ($N=100$ and $N=1000$).
Furthermore, the inset of Fig.~\ref{FreeE} (a) shows that the
theoretical result even matches well with the exact enumeration
result for small size $N=20$. As the variance parameter $g$ of
weights increases, the free energy density also decreases
(Fig.~\ref{FreeE} (b)). The same property also holds when the
variance $v$ of external field increases (the inset of
Fig.~\ref{FreeE} (b)). In the explored range of $g$ (or $v$) and
$\alpha$, Eq.~(\ref{bp3}) converges in a few steps to single fixed
point on which the free energy is calculated. Therefore, the Bethe
approximation provides an accurate estimation of free energy much
faster than other sampling based procedures which are typically slow
to reach an equilibrium state.

\begin{figure}
\centering
    \includegraphics[bb=68 19 733 532,scale=0.35]{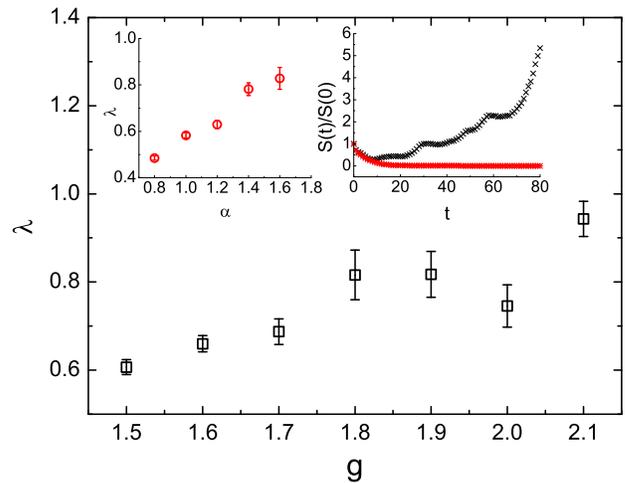}
  \caption{
  (Color online) Stability parameter $\lambda$ as a function of
  model parameters. $N=1000$, $\alpha=0.5$, and $v=0.05$. The error bars give statistical errors
      across ten random instances. The left
  inset gives stability versus $\alpha$ with $g=1.0$ and $v=0.05$.
  The right inset gives two examples (black and red) taken from the main figure at
  $g=2.1$. $S(0)$ is the initial total variance.
  }\label{sta}
\end{figure}

The stability of Eq.~(\ref{bp3}) can also be studied on {\em single
instances}. Apart from the cavity magnetization, we introduce its
variance as an extra message denoted by $\mathcal {V}_{i\rightarrow
a}$~\cite{Kaba-2003a,Montanari-2004}. The evolution of $\mathcal
{V}_{i\rightarrow a}$ follows:
\begin{equation}\label{var}
\begin{split}
    \mathcal {V}_{i\rightarrow a}&=\frac{(1-m^{2}_{i\rightarrow a})^{2}}{4}\sum_{b\in\partial i\backslash a}
    \mathcal{P}_{b\rightarrow i}\\
    &\times \left[\tanh(\Gamma_{b\rightarrow
i})-\tanh(\Gamma_{b\rightarrow i}-2w_{bi})\right]^{2},
\end{split}
\end{equation}
where $\Gamma_{b\rightarrow i}\equiv h_b+G_{b\rightarrow i}+w_{bi}$
and $\mathcal {P}_{b\rightarrow i}\equiv\sum_{j\in\partial
b\backslash i}w_{bj}^{2}\mathcal {V}_{j\rightarrow b}$. The
stability is measured by the total variance
$S(t)=\sum_{(i,a)}\mathcal {V}_{i\rightarrow a}(t)$ summed over all
connected pairs $(i,a)$, where $t$ is the iteration step. The
explosion of $S(t)$ indicates the instability of the message passing
equations, which is related to the divergence of the (non-linear)
spin glass susceptibility~\cite{Mezard-2004}, and thus the Bethe
approximation becomes inconsistent. We study this effect on single
instances of RBM as shown in Fig.~\ref{sta}. A relative strength is
denoted as $\lambda=S(t+1)/S(t)$ where $t$ denotes the step at which
the iteration converges or exceeds a prefixed maximal number
($t_{{\rm max}}=500$). $\lambda$ grows with $\alpha$ and $g$, and
the fluctuation across instances becomes strong near to the critical
point ($\lambda=1$). Note that increasing $\alpha$ has an equivalent
effect of increasing $g$. In the right inset, two typical examples
are shown. Near to the critical point, some instances have decaying
variance strength ($\lambda<1$), while some have growing strength
($\lambda>1$). Eq.~(\ref{var}) thus tells us how stable (unstable)
the iteration of Eq.~(\ref{bp3}) is for a particular RBM. The
algorithm converges in a few iteration steps to a solution unless
the recursive process is close to the instability boundary.


RBM can be used to model the real data, and the parameters are
fitted to maximize the probability of observing the training
data~\cite{Bengio-2013}. This would lead to computation of the
following quantities, $m_i\equiv \left<\sigma_i\right>$,
$\hat{m}_a\equiv\left<\tanh(\boldsymbol{w}_a\cdot\boldsymbol{\sigma}+h_a)\right>$
and
$C_{aj}\equiv\left<\tanh(\boldsymbol{w}_a\cdot\boldsymbol{\sigma}+h_a)\sigma_j\right>$,
where the average, $\langle\cdot\rangle$, is taken over the
distribution defined in Eq.~(\ref{Pobs}), which is intractable
without approximations. Here, $m_i$ is the average of visible state
$\sigma_i$, $\hat{m}_a$ is the average of hidden state $s_a$, and
$C_{aj}$ is the correlation between $s_a$ and $\sigma_i$. Although
an accurate evaluation of the above quantities requires a
sufficiently long Gibbs sampling (so-called the most difficult
negative phase in machine learning community~\cite{Bengio-2013}), we
can compute them by the message passing equations and then compare
the results with those obtained by Gibbs sampling to check the
consistency of the theory.

Following the same spirit, our theory gives the theoretical
evaluation of the above quantities as
\begin{subequations}\label{grad}
\begin{align}
m_{i}&=\tanh\left(\phi_i+\sum_{b\in\partial i}u_{b\rightarrow i}\right),\\
\hat{m}_{a}&=\int {\rm D}x\tanh(\tilde{\Xi}_{a}x+\tilde{G}_a),\\
C_{aj}&\simeq\hat{m}_{a}m_j+w_{aj}(1-m^2_j)A_{a},
\end{align}
\end{subequations}
where ${\rm D}x\equiv e^{-x^2/2}/\sqrt{2\pi}{\rm d}x$ is a Gaussian
measure. $A_a\equiv1-\int {\rm D}x\tanh^2(\tilde{\Xi}_{a}x+\tilde{G}_a)$, $\tilde{G}_a=\sum_{k\in\partial a}w_{ak}m_{k}+h_a$, and
$\tilde{\Xi}^{2}_{a}\simeq\sum_{k\in\partial a}w^{2}_{ak}(1-m^2_k)$~\cite{Huang-2014}. Eq.~(\ref{grad})
is computed based on the fixed point of the iterative algorithm
~(Eq.~(\ref{bp3})).

\begin{figure}
          \includegraphics[bb=73 14 734 540,scale=0.35]{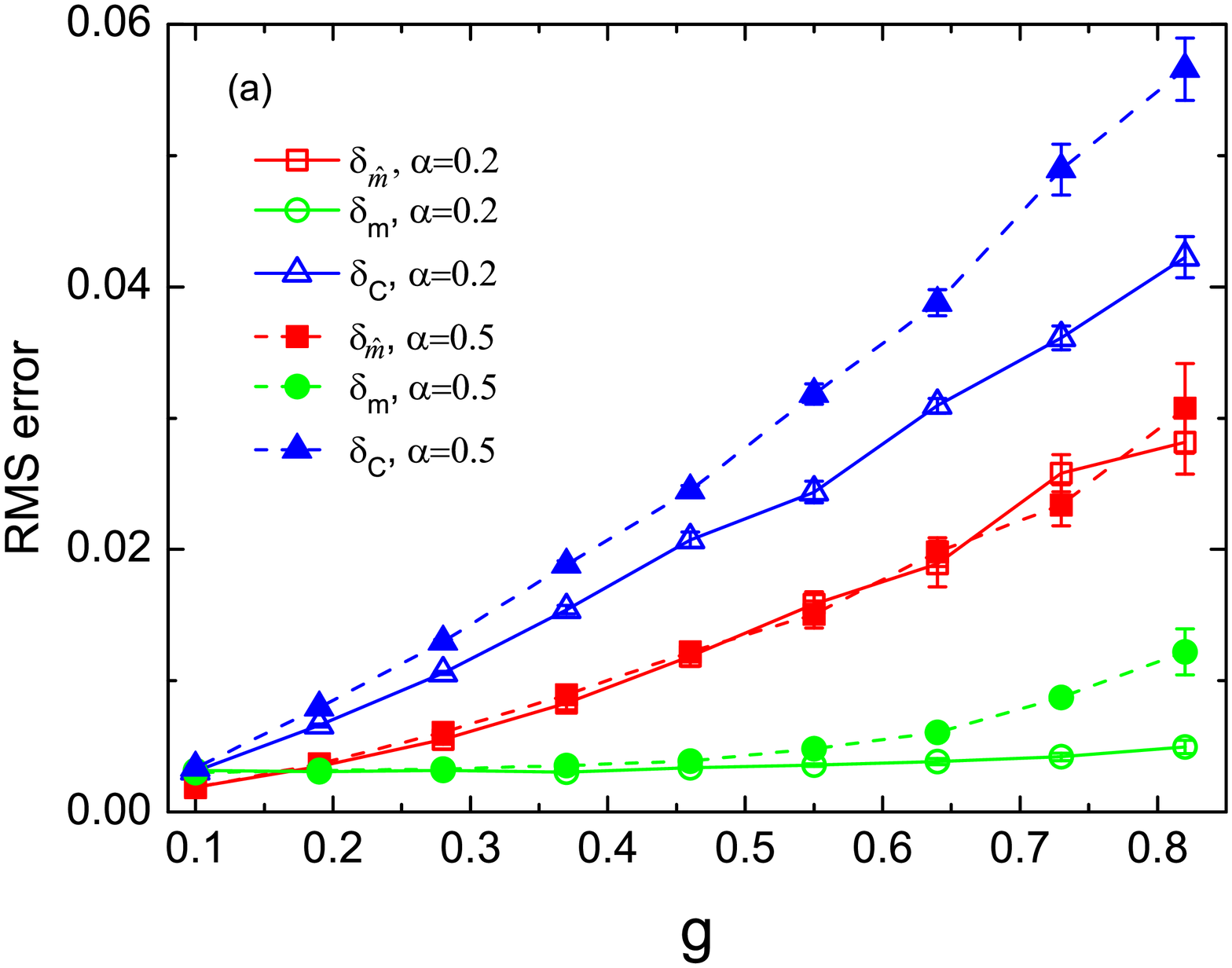}
     \vskip .1cm
     \includegraphics[bb=64 26 734 534,scale=0.35]{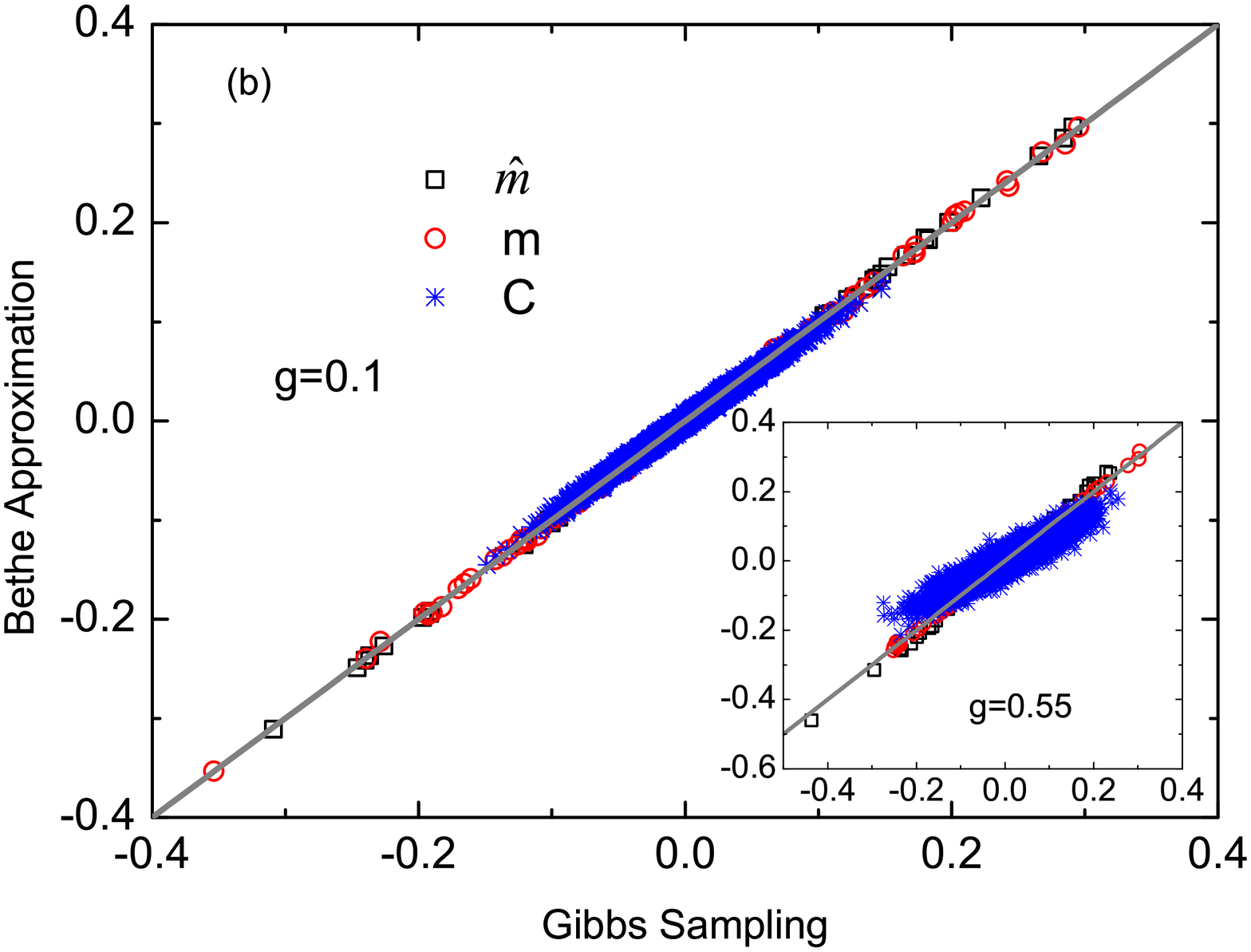}
 \vskip .1cm
 \includegraphics[bb=43 13 741 543,scale=0.35]{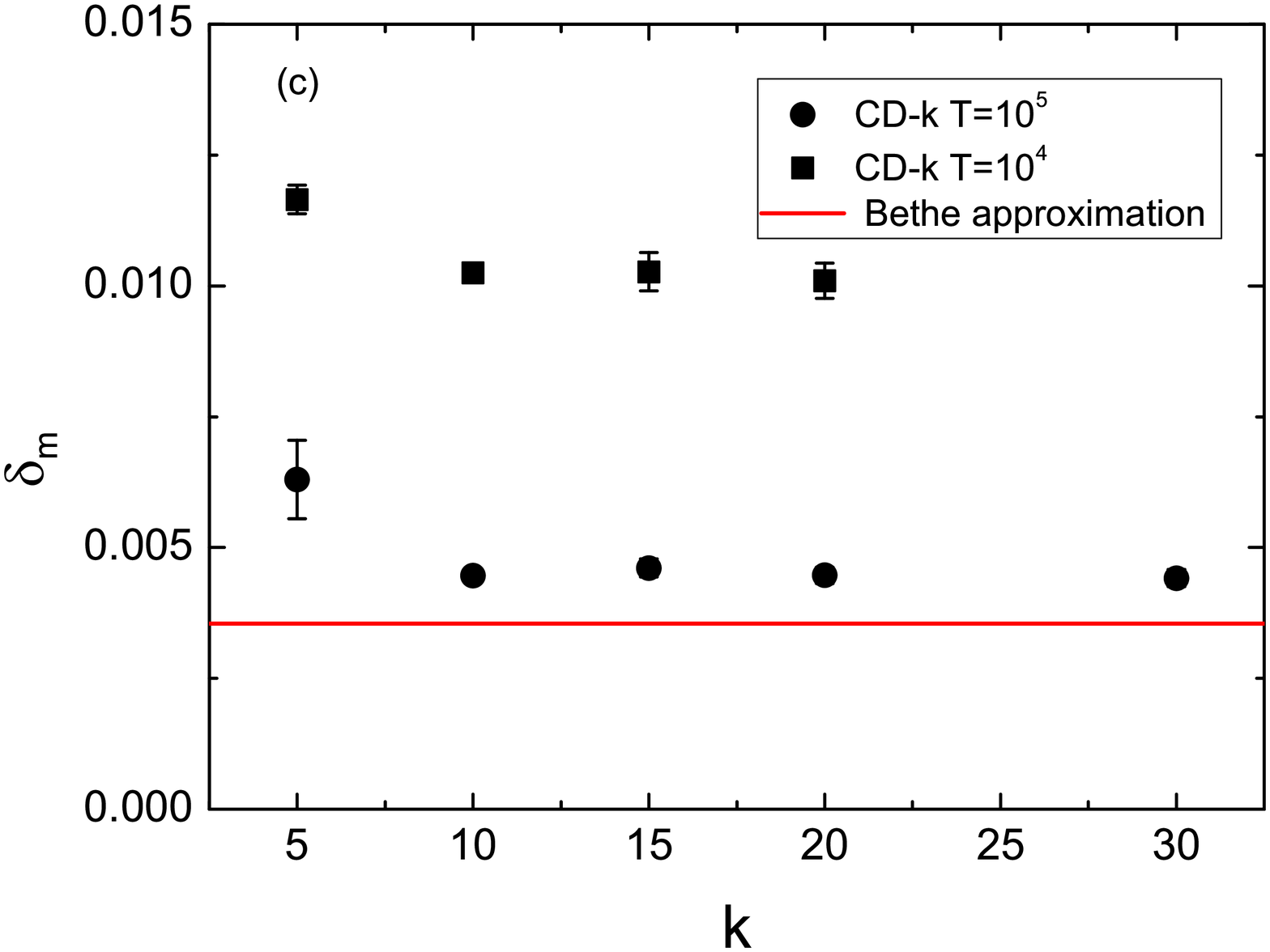}
 \vskip .1cm
  \caption{(Color online)
     Evaluation performance of mean field theory in comparison with the Gibbs sampling. Iterations of the
     recursive
      equations always converged to produce the data points. The error bars give statistical errors
      across ten random instances. (a) RMS error as a function of $g$. $N=100$, and $v=0.02$.
      (b) Scatter plot for a typical example of $N=100$, $\alpha=0.5$, $g=0.1$, and $v=0.02$. The inset shows
      an example of $g=0.55$ (other parameters do not change). The line indicates equality.
      (c) RMS error $\delta_{\boldsymbol{m}}$ reached by CD-$k$ as a function of
$k$ in comparison to the Bethe approximation. $N=100$, $\alpha=0.2$,
$g=0.55$, and $v=0.02$. The result is averaged over five random
instances.}\label{Perf}
 \end{figure}

We used alternating Gibbs sampling~\cite{Hinton-2006b} to evaluate
the equilibrium average of the gradients. More precisely, the hidden
nodes are updated in parallel according to
$P(s_a=1|\boldsymbol{\sigma})=e^{\boldsymbol{w}_a\cdot\boldsymbol{\sigma}+h_a}/(2\cosh(\boldsymbol{w}_a\cdot\boldsymbol{\sigma}+h_a))$,
while the visible nodes are then all updated in parallel according
to $P(\sigma_i=1|\boldsymbol{s})=e^{\boldsymbol{w}^{{\rm
T}}_i\cdot\boldsymbol{s}+\phi_i}/(2\cosh(\boldsymbol{w}^{{\rm
T}}_i\cdot\boldsymbol{s}+\phi_i))$ where $\boldsymbol{w}_i$ is the
$i$-th column of the weight matrix. Note that the visible nodes are
conditionally independent given the hidden states and vise
versa~\cite{Hinton-2006b}. These two steps of updates form one full
step of the alternating Gibbs sampling. If this Markov chain is run
for a sufficiently long time, the stationary (equilibrium)
distribution is expected to be reached, from which the averages can
be estimated. We test our theory in a system with $N=100$ visible
nodes and run the Markov chain with $10^6$ steps for thermal
equilibration and the other $4\times 10^6$ steps to collect a total
number of $10^5$ samples to calculate the average. We measure the
performance by the root-mean-square (RMS) error between the Gibbs
sampling (GS) result and the Bethe approximation (BA) result, which
is shown in Fig.~\ref{Perf}. The RMS error is defined as
$\delta_{\boldsymbol{Y}}\equiv
\sqrt{\frac{1}{|\boldsymbol{Y}|}\sum_{i=1}^{|\boldsymbol{Y}|}\bigl(Y_{i}^{{\rm
GS}}-Y_{i}^{{\rm BA}}\bigr)^{2}}$ where $\boldsymbol{Y}$ takes
either $\boldsymbol{m}$, $\boldsymbol{\hat{m}}$, or
$\boldsymbol{C}$, and $|\boldsymbol{Y}|$ indicates the number of
these parameters. Small RMS error indicates that the Bethe
approximation is accurate.

As shown in Fig.~\ref{Perf} (a), all evaluation errors grow with the
weight strength $g$, which is reasonable since our mean field theory
will break down when the network enters a strongly correlated state,
as already shown by the stability analysis. In a similar manner, the
error grows with the hidden node density $\alpha$, because each
hidden node puts a constraint to the network and all constraints
compete with each other to give an equilibrium state, resulting in
strong correlations with high $\alpha$. However, the magnitude of
all errors is small, implying that one can acquire accurate
estimation of gradients of log-likelihood function by passing
messages on a factor graph as well. We show this point more clearly
with a scatter plot in Fig.~\ref{Perf} (b) (Bethe approximation
result versus Gibbs sampling result). This accuracy is obtained by
requiring much fewer computational costs compared with the Gibbs
sampling. The comparison further confirms the efficiency of the
proposed mean field method across a wide range of model parameters.

Note that the Gibbs sampling result serves as the ground truth here,
since we run the Markov chain for a long time. More practically, one
can estimate the statistics by $k$-steps contrastive divergence
(CD-$k$) algorithm~\cite{Hinton-2006b} that requires a time
complexity of $\mathcal {O}(kTMN)$, where $T$ denotes the number of
sample particles. However, to reach a similar accuracy as the Bethe
approximation, it typically requires $k\geq10$ and $T\sim 10^{5}$
under the current setup (Fig.~\ref{Perf} (c)). In contrast, the
Bethe approximation yields a time complexity of $\mathcal {O}(nMN)$
with $n<100$, where $n$ is the number of iterations and one
iteration involves the update of $MN$ cavity messages.

In conclusion, we propose a mean field theory for the RBM, a widely
used model in machine learning community and biological data
analysis. The theory captures nearest neighbors' correlations by
operating on the cavity factor graph (by removing factor nodes),
leading to an approximate estimation of the free energy function
(log-likelihood function) for single instances of large-size
networks, for which the standard Gibbs sampling procedure becomes
prohibitively slow to get a reliable result (e.g., for evaluating
the log-likelihood function for cross-validation analysis).
Moreover, we replace the normal Gibbs sampling with a mean field
computation based on message passing algorithm, to estimate the
gradients of log-likelihood function and show its efficiency by
extensive numerical simulations on single instances. The nature of
this fast inference lies in the fact that, the information is
exchanged locally between factor nodes and visible nodes, to reach a
coherent fixed point, which may provide a computation paradigm for
probabilistic inference in neural networks. We expect the mean field
theory inspired calculation will be useful in practical applications
and bring more insights to understand the RBM and its role in deep
learning~\cite{Mehta-2014}.


This work was supported by
RIKEN Brain Science Institute and the Brain Mapping by Integrated Neurotechnologies for Disease
Studies (Brain/MINDS) by the Ministry of Education, Culture, Sports,
Science and Technology of Japan (MEXT).

\begin{thebibliography}{18}
\expandafter\ifx\csname
natexlab\endcsname\relax\def\natexlab#1{#1}\fi
\expandafter\ifx\csname bibnamefont\endcsname\relax
  \def\bibnamefont#1{#1}\fi
\expandafter\ifx\csname bibfnamefont\endcsname\relax
  \def\bibfnamefont#1{#1}\fi
\expandafter\ifx\csname citenamefont\endcsname\relax
  \def\citenamefont#1{#1}\fi
\expandafter\ifx\csname url\endcsname\relax
  \def\url#1{\texttt{#1}}\fi
\expandafter\ifx\csname urlprefix\endcsname\relax\def\urlprefix{URL
}\fi \providecommand{\bibinfo}[2]{#2}
\providecommand{\eprint}[2][]{\url{#2}}

\bibitem[{\citenamefont{Hinton and Salakhutdinov}(2006)}]{Hinton-2006a}
\bibinfo{author}{\bibfnamefont{G.~E.} \bibnamefont{Hinton}} \bibnamefont{and}
  \bibinfo{author}{\bibfnamefont{R.~R.} \bibnamefont{Salakhutdinov}},
  \bibinfo{journal}{Science} \textbf{\bibinfo{volume}{313}},
  \bibinfo{pages}{504} (\bibinfo{year}{2006}).

\bibitem[{\citenamefont{Bengio et~al.}(2013)\citenamefont{Bengio, Courville,
  and Vincent}}]{Bengio-2013}
\bibinfo{author}{\bibfnamefont{Y.}~\bibnamefont{Bengio}},
  \bibinfo{author}{\bibfnamefont{A.}~\bibnamefont{Courville}},
  \bibnamefont{and} \bibinfo{author}{\bibfnamefont{P.}~\bibnamefont{Vincent}},
  \bibinfo{journal}{Pattern Analysis and Machine Intelligence, IEEE
  Transactions on} \textbf{\bibinfo{volume}{35}}, \bibinfo{pages}{1798}
  (\bibinfo{year}{2013}).

\bibitem[{\citenamefont{K\"oster et~al.}(2014)\citenamefont{K\"oster,
  Sohl-Dickstein, Gray, and Olshausen}}]{koster-2014}
\bibinfo{author}{\bibfnamefont{U.}~\bibnamefont{K\"oster}},
  \bibinfo{author}{\bibfnamefont{J.}~\bibnamefont{Sohl-Dickstein}},
  \bibinfo{author}{\bibfnamefont{C.~M.} \bibnamefont{Gray}}, \bibnamefont{and}
  \bibinfo{author}{\bibfnamefont{B.~A.} \bibnamefont{Olshausen}},
  \bibinfo{journal}{PLoS Comput Biol} \textbf{\bibinfo{volume}{10}},
  \bibinfo{pages}{e1003684} (\bibinfo{year}{2014}).

\bibitem[{\citenamefont{Hinton}(2002)}]{Hinton-2002}
\bibinfo{author}{\bibfnamefont{G.}~\bibnamefont{Hinton}},
  \bibinfo{journal}{Neural Computation} \textbf{\bibinfo{volume}{14}},
  \bibinfo{pages}{1771} (\bibinfo{year}{2002}).

\bibitem[{\citenamefont{Hinton et~al.}(2006)\citenamefont{Hinton, Osindero, and
  Teh}}]{Hinton-2006b}
\bibinfo{author}{\bibfnamefont{G.}~\bibnamefont{Hinton}},
  \bibinfo{author}{\bibfnamefont{S.}~\bibnamefont{Osindero}}, \bibnamefont{and}
  \bibinfo{author}{\bibfnamefont{Y.}~\bibnamefont{Teh}},
  \bibinfo{journal}{Neural Computation} \textbf{\bibinfo{volume}{18}},
  \bibinfo{pages}{1527} (\bibinfo{year}{2006}).

\bibitem[{\citenamefont{M\'ezard and Parisi}(2001)}]{cavity-2001}
\bibinfo{author}{\bibfnamefont{M.}~\bibnamefont{M\'ezard}} \bibnamefont{and}
  \bibinfo{author}{\bibfnamefont{G.}~\bibnamefont{Parisi}},
  \bibinfo{journal}{Eur. Phys. J. B} \textbf{\bibinfo{volume}{20}},
  \bibinfo{pages}{217} (\bibinfo{year}{2001}).

\bibitem[{\citenamefont{Smolensky}(1986)}]{Smolensky-1986}
\bibinfo{author}{\bibfnamefont{P.}~\bibnamefont{Smolensky}}
  (\bibinfo{publisher}{MIT Press}, \bibinfo{address}{Cambridge, MA, USA},
  \bibinfo{year}{1986}), chap. \bibinfo{chapter}{Information Processing in
  Dynamical Systems: Foundations of Harmony Theory}, pp.
  \bibinfo{pages}{194--281}.

\bibitem[{\citenamefont{Freund and Haussler}(1994)}]{Freund-1994}
\bibinfo{author}{\bibfnamefont{Y.}~\bibnamefont{Freund}} \bibnamefont{and}
  \bibinfo{author}{\bibfnamefont{D.}~\bibnamefont{Haussler}},
  \bibinfo{type}{Tech. Rep.}, \bibinfo{address}{Santa Cruz, CA, USA}
  (\bibinfo{year}{1994}).

\bibitem[{\citenamefont{Kschischang et~al.}(2001)\citenamefont{Kschischang,
  Frey, and Loeliger}}]{Frey-2001}
\bibinfo{author}{\bibfnamefont{F.~R.} \bibnamefont{Kschischang}},
  \bibinfo{author}{\bibfnamefont{B.~J.} \bibnamefont{Frey}}, \bibnamefont{and}
  \bibinfo{author}{\bibfnamefont{H.-A.} \bibnamefont{Loeliger}},
  \bibinfo{journal}{IEEE Trans. Inf. Theory} \textbf{\bibinfo{volume}{47}},
  \bibinfo{pages}{498} (\bibinfo{year}{2001}).

\bibitem[{\citenamefont{M\'ezard and Montanari}(2009)}]{MM-2009}
\bibinfo{author}{\bibfnamefont{M.}~\bibnamefont{M\'ezard}} \bibnamefont{and}
  \bibinfo{author}{\bibfnamefont{A.}~\bibnamefont{Montanari}},
  \emph{\bibinfo{title}{Information, Physics, and Computation}}
  (\bibinfo{publisher}{Oxford University Press}, \bibinfo{address}{Oxford},
  \bibinfo{year}{2009}).

\bibitem[{\citenamefont{Huang et~al.}(2013)\citenamefont{Huang, Wong, and
  Kabashima}}]{Huang-JPA2013}
\bibinfo{author}{\bibfnamefont{H.}~\bibnamefont{Huang}},
  \bibinfo{author}{\bibfnamefont{K.~Y.~M.} \bibnamefont{Wong}},
  \bibnamefont{and}
  \bibinfo{author}{\bibfnamefont{Y.}~\bibnamefont{Kabashima}},
  \bibinfo{journal}{J. Phys. A: Math. Theor.} \textbf{\bibinfo{volume}{46}},
  \bibinfo{pages}{375002} (\bibinfo{year}{2013}).

\bibitem[{\citenamefont{Kappen and Rodriguez}(1998)}]{Kappen-1998}
\bibinfo{author}{\bibfnamefont{H.~J.} \bibnamefont{Kappen}} \bibnamefont{and}
  \bibinfo{author}{\bibfnamefont{F.~B.} \bibnamefont{Rodriguez}},
  \bibinfo{journal}{Neural Comput} \textbf{\bibinfo{volume}{10}},
  \bibinfo{pages}{1137} (\bibinfo{year}{1998}).

\bibitem[{\citenamefont{Heskes}(2004)}]{Heskes-2004}
\bibinfo{author}{\bibfnamefont{T.}~\bibnamefont{Heskes}},
  \bibinfo{journal}{Neural Comput} \textbf{\bibinfo{volume}{16}},
  \bibinfo{pages}{2379} (\bibinfo{year}{2004}).

\bibitem[{\citenamefont{Kabashima}(2003)}]{Kaba-2003a}
\bibinfo{author}{\bibfnamefont{Y.}~\bibnamefont{Kabashima}},
  \bibinfo{journal}{J. Phys. A} \textbf{\bibinfo{volume}{36}},
  \bibinfo{pages}{11111} (\bibinfo{year}{2003}).

\bibitem[{\citenamefont{Montanari and Ricci-Tersenghi}(2004)}]{Montanari-2004}
\bibinfo{author}{\bibfnamefont{A.}~\bibnamefont{Montanari}} \bibnamefont{and}
  \bibinfo{author}{\bibfnamefont{F.}~\bibnamefont{Ricci-Tersenghi}},
  \bibinfo{journal}{Phys. Rev. B} \textbf{\bibinfo{volume}{70}},
  \bibinfo{pages}{134406} (\bibinfo{year}{2004}).

\bibitem[{\citenamefont{Rivoire et~al.}(2004)\citenamefont{Rivoire, Biroli,
  Martin, and M\'ezard}}]{Mezard-2004}
\bibinfo{author}{\bibfnamefont{O.}~\bibnamefont{Rivoire}},
  \bibinfo{author}{\bibfnamefont{G.}~\bibnamefont{Biroli}},
  \bibinfo{author}{\bibfnamefont{O.}~\bibnamefont{Martin}}, \bibnamefont{and}
  \bibinfo{author}{\bibfnamefont{M.}~\bibnamefont{M\'ezard}},
  \bibinfo{journal}{Eur. Phys. J. B} \textbf{\bibinfo{volume}{37}},
  \bibinfo{pages}{55} (\bibinfo{year}{2004}).

\bibitem[{\citenamefont{Huang and Kabashima}(2014)}]{Huang-2014}
\bibinfo{author}{\bibfnamefont{H.}~\bibnamefont{Huang}} \bibnamefont{and}
  \bibinfo{author}{\bibfnamefont{Y.}~\bibnamefont{Kabashima}},
  \bibinfo{journal}{J. Stat. Mech.: Theory Exp} p. \bibinfo{pages}{P05020}
  (\bibinfo{year}{2014}).

\bibitem[{\citenamefont{Mehta and Schwab}(2014)}]{Mehta-2014}
\bibinfo{author}{\bibfnamefont{P.}~\bibnamefont{Mehta}} \bibnamefont{and}
  \bibinfo{author}{\bibfnamefont{D.~J.} \bibnamefont{Schwab}},
  \bibinfo{journal}{ArXiv e-prints}  (\bibinfo{year}{2014}),
  \eprint{1410.3831}.

\end{thebibliography}


\end{document}